\documentclass[twocolumn,aps,prl,amsmath,showpacs,amssymb,floatfix]{revtex4}
\usepackage{tabularx}
\usepackage{bm}
\usepackage{subfigure}
\usepackage{euscript}
\usepackage{graphicx}
\usepackage{color}
\usepackage[colorlinks=true,linkcolor=blue]{hyperref}%
\usepackage{amsfonts}
\usepackage{exscale}
\usepackage{amsbsy}

\begin{document}

\title{Measuring the Spin Polarization of a Ferromagnet: an Application of \\
Time-Reversal Invariant Topological Superconductor}

\author{Zhongbo Yan}
\author{Shaolong Wan}
\email[]{slwan@ustc.edu.cn}
\affiliation{Institute for Theoretical
Physics and Department of Modern Physics University of Science and
Technology of China, Hefei, 230026, P. R. China}
\date{\today}

\begin{abstract}
The spin polarization (SP) of the ferromagnet (FM) is a quantity of fundamental importance in spintronics.
In this work, we propose a quasi-one-dimensional junction structure composed of a FM and a time-reversal
invariant topological superconductor (TRITS) with un-spin-polarized pairing type to determine the SP of
the FM.  We find that due to the topological property of the TRITS, the zero-bias conductance (ZBC) of
the FM/TRITS junction which is directly related to the SP is a non-quantized but topological quantity.
The ZBC only depends on the parameters of the FM, it is independent of the interface scattering potential
and the Fermi surface mismatch between the FM and the superconductor, and is robust against to the magnetic
proximity effect, therefore, compared to the traditional FM/$s$-wave superconductor junction, the topological
property of the ZBC makes this setup a much more direct and simplified way to determine the SP.
\end{abstract}

\pacs{03.65.Vf, 71.10.Pm, 74.50.+r, 73.40.-c, 85.75.-d}

\maketitle

{\it Introduction---} Because of hosting exotic non-Abelian zero modes
\cite{E. Majorana, D. A. Ivanov} which have great potential in topological
quantum computation (TQC) \cite{Kitaev, S. Das Sarma, C. Nayak}, topological
superconductors (TSs) in every dimension have raised strong and lasting interests
for more than a decade \cite{N. Read, A. Kitaev, G. E. Volovik, Liang Fu, C. W. Zhang,
M. Sato, X.-L. Qi, Y. Tanaka, J. Linder, Jay D. Sau, Roman M. Lutchyn, Y. Oreg,
J. Alicea1, A. C. Potter, S. Sasaki, S. Nakosai, A. Keselman}. Due to the
nontrivial topology of the energy bands, TSs have many properties that are
fundamentally different from the normal superconductors (NSs). One of the
most remarkable difference is that the zero-bias conductance (ZBC) of a
normal metal (NM)/TS junction is a quantized quantity of topological
nature \cite{A. R. Akhmerov, K. T. Law, K. Flensberg}, while for a NM/NS
junction, the ZBC is parameter-dependent and can be greatly suppressed by
the interface scattering potential \cite{G. E. Blonder}.

As ferromagnet (FM) plays a crucial role in spintronics, the spin polarization
(SP) of the FM is of fundamental importance \cite{S. A. Wolf, I. Zutic}.  To
determine the SP, a general approach is to detect the tunneling spectroscopy of
the FM/$s$-wave superconductor junction \cite{R. J. Soulen, S. K. Upadhyay, Y. Ji,
G. J. Strijkers}. The underlying mechanism is based on the fact that for a ballistic
NM/$s$-wave superconductor junction, an electron with Fermi energy injected from the
NM to the superconductor will be completely reflected as a hole with opposite spin,
which is known as spin-opposite Andreev reflection \cite{A. F. Andreev}, however, when
the metal is a FM, due to the mismatch of the Fermi surface between the two spin
degrees, some of the majority spin electrons can not undergo the spin-opposite
Andreev reflection \cite{M. J. M. de Jong}, instead, they are reflected as themselves,
which is known as normal reflection, consequently, compared to the NM/$s$-wave
superconductor junction, the conductance of the FM/$s$-wave superconductor junction
is decreased, and the decrement monotonically increases with the mismatch increasing.
As a result, the SP can be quantitatively determined by the tunneling spectroscopy.

Although the idea of the above mechanism is generally applied for every FM/$s$-wave
superconductor junction, the concrete decrement can also be induced by other factors,
such as the interface scattering potential and the Fermi surface mismatch between the
FM and the superconductor \cite{Elina Tuuli}. As a result, for general FM/$s$-wave
superconductor junction, the tunneling spectroscopy may involve many parameters, and
consequently, the SP is very hard to be precisely resolved from the tunneling spectroscopy
\cite{I. I. Mazin1, T. Y. Chen}. However, in this work, we find that if the normal $s$-wave
superconductor is substituted by a time-reversal invariant (TRI) TS with un-spin-polarized
pairing type, then as the ZBC turns out to be a topological quantity only related to the
parameters of the FM, the process to determine the SP becomes much more direct and simplified.

{\it Theoretical model---} So far, the greatest experimental progress made on the
transport study of TS is in one dimension \cite{V. Mourik, A. Das, M. T. Deng,
A. D. K. Finck, L. P. Rokhinson, S. Nadj-Perge}. For generality, in this work we
consider the FM is a quasi-one-dimensional wire, with length $L$ in $x$-direction
and width $W$ in $y$-direction, and $L>>W$. Correspondingly, for the TS, the width
is also given by $W$, and the length is assumed to be infinite for simplicity. Then
the Hamiltonian describing the junction under the representation $\hat{\Psi}^{\dag}(x,y)
=(\hat{\psi}^{\dag}_{\uparrow}(x,y),\hat{\psi}_{\downarrow}(x,y),\hat{\psi}^{\dag}_{\downarrow}(x,y),
\hat{\psi}_{\uparrow}(x,y))$ is given by
\begin{eqnarray}
\mathcal{H}=\tau_{z}\left[-\frac{\hbar^{2}}{2m}(\partial^{2}_{x}+\partial^{2}_{y}) - \mu(x,y) + V(x,y)
\right] + \tau_{x}\Delta(x,y), \label{1}
\end{eqnarray}
where $\vec{\tau}=(\tau_{x}, \tau_{y}, \tau_{z})$ are Pauli matrices
in particle-hole space, $V(x,y)$ is potential induced by disorder, external
field, $etc$, here we assume it takes the form $-M\tau_{z}\sigma_{z}\Theta(-x)
+V\delta(x)$, the former term denotes the magnetization of the FM, $\sigma_{z}$
is a Pauli matrix acting on the spin space, $\Theta(-x)$ is the Heaviside function,
the latter term denotes the scattering potential at the interface. $\mu(x,y)$ is
the chemical potential, we set $\mu(x,y)=\mu_{f}$ (or $E_{F}$) for the ferromagnetic
part ($x<0$) and $\mu(x,y) = \mu_{s}$ for the superconductor ($x>0$). $\Delta(x,y)=
-i\Delta_{0}\Theta(x)\partial_{x}$ is the pairing potential, which is assumed to be
$p$-wave type and homogeneous at $x>0$ and vanish at $x<0$ for the sake of theoretical
simplicity. The mass $m$ of the particle is assumed to be positive and the same throughout
the system.

As the system is strongly confined in $y$-direction, the system will form a series of subbands
with band index $n$ a good quantum number. Then the field operator can be expressed as $\hat{\psi}_{\sigma}(x,y)=\sum_{n}\hat{\psi}_{n\sigma}(x)\chi_{n}(y)$, where $\chi_{n}(y)=\sqrt
{\frac{2}{W}}\sin(k_{n}y)$, with $k_{n}=n\pi/W$. By a Fourier transformation $\hat{\psi}_{n\sigma}(x)
=\int\frac{dk}{2\pi}e^{ikx} \hat{c}_{n\sigma,k}$, the Hamiltonians for the ferromagnetic part and
the superconducting part under the representation $\hat{\Psi}^{\dag}_{nk}=(\hat{c}_{n\uparrow,k}^{\dag},
\hat{c}_{n\downarrow,-k},\hat{c}_{n\downarrow,k}^{\dag},\hat{c}_{n\uparrow,-k})$ are given as
\begin{eqnarray}
&&\mathcal{H}_{F}(k)=[\frac{\hbar^{2}k^{2}}{2m}
+\epsilon_{n}-\mu_{f}]\tau_{z}-M\tau_{0}\sigma_{z},\nonumber\\
&&\mathcal{H}_{S}(k)=[\frac{\hbar^{2}k^{2}}{2m}
+\epsilon_{n}-\mu_{s}]\tau_{z}+\Delta(k)\tau_{x}, \label{2}
\end{eqnarray}
respectively, where $\epsilon_{n}=n^{2}\hbar^{2}\pi^{2}/(2mW^{2})$, $\Delta(k)=\Delta_{0}k$.
As $\tau_{z}\mathcal{H}_{S}(k)\tau_{z}=\mathcal{H}_{S}^{*}(-k)$, $\tau_{x}\mathcal{H}_{S}(k)\tau_{x}=
-\mathcal{H}_{S}^{*}(-k)$, $\tau_{y}\mathcal{H}_{S}(k)\tau_{y}=-\mathcal{H}_{S}(k)$, and
$(\tau_{z}K)^{2}=1$, $\mathcal{H}_{S}(k)$ belongs to the BDI class \cite{A. P. Schnyder, A. Y. Kitaev}.
From Eq.(\ref{2}), it is direct to obtain the excitation energy spectra of the FM,
$E_{fn\uparrow(\downarrow)}=\hbar^{2}k^{2}/2m+\epsilon_{n}-\mu_{f}\mp M$, then the particle number partition $(N_{\uparrow},N_{\downarrow})$ can be directly obtained as $N_{\uparrow}=(L/\pi)\sum_{n}^{'}\sqrt{2m(\mu_{f}
+M-\epsilon_{n})}$, $N_{\downarrow}=(L/\pi)\sum_{n}^{''}\sqrt{2m(\mu_{f}-M-\epsilon_{n})}$, where the
two superscripts mean that the two summations are limited by two upper limit $n^{'}$ and $n^{''}$,
respectively. $n^{'}$ satisfies $(\mu_{f}+M-\epsilon_{n^{'}})>0$ and $(\mu_{f}+M-\epsilon_{n^{'}+1})<0$,
similarly $n^{''}$ satisfies $(\mu_{f}-M-\epsilon_{n^{''}})>0$ and $(\mu_{f}-M-\epsilon_{n^{''}+1})<0$.
If the particle number partition $(N_{\uparrow},N_{\downarrow})$ is known, the SP, which is
defined as \cite{I. I. Mazin2}

\begin{eqnarray}
P&\equiv&\frac{\sum_{n}^{'}N_{n\uparrow}(E_{F})v_{F,n\uparrow}^{2}-\sum_{n}^{''}N_{n\downarrow}(E_{F})v_{F,n\downarrow}^{2}}
{\sum_{n}^{'}N_{n\uparrow}(E_{F})v_{F,n\uparrow}^{2}+\sum_{n}^{''}N_{n\downarrow}(E_{F})v_{F,n\downarrow}^{2}}\nonumber\\
&=&\frac{N_{\uparrow}-N_{\downarrow}}{N_{\uparrow}+N_{\downarrow}},\label{3}
\end{eqnarray}
where $N_{n\sigma}(E_{F})$ is the density of states at the Fermi energy and $v_{F,n\sigma}$
is the Fermi velocity, can be directly obtained. Note that in the quasi-one-dimensional case,
as $N_{n\uparrow}(E_{F})v_{F,n\uparrow}=N_{n\downarrow}(E_{F})v_{F,n\downarrow}=const$, the
ballistic definition $P=(N_{\uparrow}(E_{F})v_{F\uparrow}-N_{\downarrow}(E_{F})v_{F\downarrow})/
(N_{\uparrow}(E_{F})v_{F\uparrow}+N_{\downarrow}(E_{F})v_{F\downarrow})$ does not apply
\cite{R. J. Soulen}. For the superconducting part, the quasi-particle energy spectra
is given as $E_{sn}=\sqrt{(\hbar^{2}k^{2}/2m+\epsilon_{n}-\mu_{s})^{2}+\Delta_{0}^{2}k^{2}}$.
When $\mu_{s}-\epsilon_{n}>0$, the bands with index smaller than $n+1$ are all of nontrivial
topology \cite{B. A. Bernevig}. In this work, we first consider $\epsilon_{1}<\mu_{s}<\epsilon_{2}$,
in other words, only bands with index $n=1$ are of nontrivial topology.

{\it Relation between ZBC and SP---} Due to the orthogonality
of $\{\chi_{n}(y)\}$, if an electron with spin-up, excitation energy $E$ and band index $n$
is injected from the FM, the wave function in the FM is given as $\psi_{f,n}(x<0)=\vec{e}_{1}e^{iq_{n\uparrow,e}x}+b_{n\uparrow}\vec{e}_{1}e^{-iq_{n\uparrow,e}x}+
a_{n\downarrow}\vec{e}_{2}e^{iq_{n\downarrow,h}x}+b_{n\downarrow}\vec{e}_{3}e^{-iq_{n\downarrow,e}x}
+a_{n\uparrow}\vec{e}_{4}e^{iq_{n\uparrow,h}x}$, where $\vec{e}_{1}=(1,0,0,0)^{T}$, $\vec{e}_{2}=(0,1,0,0)^{T}$£¬
$\vec{e}_{3}=(0,0,1,0)^{T}$, and $\vec{e}_{4}=(0,0,0,1)^{T}$.
$q_{n\uparrow,e}(E)=\sqrt{2m(\mu_{f}+M+E-\epsilon_{n})}$, $q_{n\downarrow,h}(E)=\sqrt{2m(\mu_{f}-M-E-\epsilon_{n})}$,
$q_{n\downarrow,e}(E)=\sqrt{2m(\mu_{f}-M+E-\epsilon_{n})}$, and $q_{n\uparrow,h}(E)=\sqrt{2m(\mu_{f}+M-E-\epsilon_{n})}$.
$b_{n\uparrow}$ and $b_{n\downarrow}$ denote the amplitudes corresponding to spin-equal and
spin-opposite normal reflection, respectively. $a_{n\uparrow}$ and $a_{n\downarrow}$ denote
the amplitudes corresponding to spin-equal and spin-opposite Andreev reflection, respectively.
In this work, we are only interested in the special case with $E=0$. When $E=0$, the wave function in the
superconductor is very simple. If $n=1$, corresponding to the band of nontrivial topology, $\psi_{s,n}(x>0)=c_{n1}
\vec{e}_{5}e^{-k_{n+}x}+d_{n1}\vec{e}_{5}e^{-k_{n-}x}+c_{n2}\vec{e}_{6}e^{-k_{n+}x}+d_{n2}\vec{e}_{6}e^{-k_{n-}x}$,
where $\vec{e}_{5}=(i,1,0,0)^{T}$ and $\vec{e}_{6}=(0,0,i,1)^{T}$. While for $n\geq2$, corresponding to the bands of
trivial topology, $\psi_{s,n}(x>0)=c_{n1}\vec{e}_{5}e^{-k_{n+}x}+d_{n1}\vec{e}_{7}e^{-k_{n-}x}+c_{n2}\vec{e}_{6}e^{-k_{n+}x}
+d_{n2}\vec{e}_{8}e^{-k_{n-}x}$, where $\vec{e}_{7}=(-i,1,0,0)^{T}$ and $\vec{e}_{8}=(0,0,-i,1)^{T}$. As $k_{n+}$
and $k_{n-}$  will not show up in the results, we do not write down their expressions explicitly here.

If the superconductor is only weak pairing which means that only when the band minimum is lower than $\mu_{s}$,
the band is metallic and has states to pair to be superconducting \cite{B. A. Bernevig}, we only need to consider
the $n=1$ bands. However, for generality, here we consider all bands are paired to be superconducting.

Again due to the orthogonality of $\{\chi_{n}(y)\}$, the boundary conditions of
the wave functions at the interface is given as \cite{Z. B. Yan}
\begin{eqnarray}
&&\psi_{f,n}(0)=\psi_{s,n}(0),\nonumber\\
&&v_{s}\psi_{s,n}(0^{+})-v_{f}\psi_{f,n}(0^{-})=-iZ\tau_{z}\sigma_{0}\psi_{s,n}(0),\label{4}
\end{eqnarray}
where $v_{s}=\partial_{k}\mathcal{H}_{S}/\hbar$, $v_{f}=\partial_{k}\mathcal{H}_{F}/\hbar$,
$Z=2V/\hbar$. Based on Eq.(\ref{4}), all coefficients can be directly obtained, and then
according to the Blonder-Tinkham-Klapwijk formula \cite{G. E. Blonder}, the ZBC
is given as
\begin{eqnarray}
G(0)=\frac{e^{2}}{h}\sum_{n_{\pm}}(1+A_{n_{\pm}\uparrow}+A_{n_{\pm}\downarrow}
-B_{n_{\pm}\uparrow}-B_{n_{\pm}\downarrow}),\label{5}
\end{eqnarray}
where $+$(-) denotes that the injected electron is spin-up (spin-down).
The summation on majority spin band number $n_{+}$ (minority spin band
number $n_{-}$) goes from $1$ to $n^{'}$ ($n^{''}$).
$A_{n_{+}\uparrow}=q_{n\uparrow,h}(0)|a_{n\uparrow}|^{2}/q_{n\uparrow,e}(0)$,
$B_{n_{+}\uparrow}=|b_{n\uparrow}|^{2}$,
$A_{n_{+}\downarrow}=q_{n\downarrow,h}(0)|a_{n\downarrow}|^{2}/q_{n\uparrow,e}(0)$,
$B_{n_{+}\downarrow}=q_{n\downarrow,e}(0)|b_{n\downarrow}|^{2}/q_{n\uparrow,e}(0)$,
$A_{n_{-}\uparrow}=q_{n\uparrow,h}(0)|a_{n\uparrow}|^{2}/q_{n\downarrow,e}(0)$,
$B_{n_{-}\downarrow}=|b_{n\downarrow}|^{2}$,
$A_{n_{-}\downarrow}=q_{n\downarrow,h}(0)|a_{n\downarrow}|^{2}/q_{n\downarrow,e}(0)$,
$B_{n_{-}\uparrow}=q_{n\uparrow,e}(0)|b_{n\uparrow}|^{2}/q_{n\downarrow,e}(0)$.
Due to the current conservation, these quantities satisfy the constraint:
$A_{n_{\pm}\uparrow}+A_{n_{\pm}\downarrow}+B_{n_{\pm}\downarrow}+B_{n_{\pm}\downarrow}=1$.
This constraint can simplify the conductance formula as
\begin{eqnarray}
G(0)=\frac{2e^{2}}{h}\sum_{n_{\pm}}(A_{n_{\pm}\uparrow}
+A_{n_{\pm}\downarrow}).\label{6}
\end{eqnarray}

Based on Eq.(\ref{4}), a direct calculation shows that $A_{n_{+}\uparrow}$
and $A_{n_{-}\downarrow}$ always vanish and \cite{SM}
\begin{eqnarray}
A_{n_{+}\downarrow}=A_{n_{-}\uparrow}=\left\{\begin{array}{cc}
\frac{4q_{n\uparrow}q_{n\downarrow}}{(q_{n\uparrow}+q_{n\downarrow})^{2}},&n=1\\0,&n\geq2.\end{array}\right. \label{7}
\end{eqnarray}
where $q_{n\uparrow(\downarrow)}=q_{n\uparrow(\downarrow),e}(0)=q_{n\uparrow(\downarrow),h}(0)$.
$A_{n_{+}\uparrow}$ and $A_{n_{-}\downarrow}$, both denoting the spin-equal Andreev reflection,
taking value zero is a natural result since the superconductor is with un-spin-polarized pairing
type. The non-vanishing quantities only depend on the parameters of the FM, they are independent of the
scattering potential and the parameters of the superconductor, which suggests that they are of
topological nature.  Substituting Eq.(\ref{7}) into Eq.(\ref{6}), it is direct to obtain
\begin{eqnarray}
\bar{G}(0)\equiv\frac{hG(0)}{e^{2}}=\frac{16q_{1\uparrow}q_{1\downarrow}}{(q_{1\uparrow}+q_{1\downarrow})^{2}}.  \label{8}
\end{eqnarray}
The zero-bias conductance is only related to the lowest spin-up and spin-down
subband of the FM. As a result, it is found that only in the strict one-dimensional
limit, $G(0)$ has enough information to directly determine the polarization of the FM.
In the strict one-dimensional limit, $i.e.$, $\epsilon_{1}<\mu_{f}$ and $\mu_{f}-\epsilon_{1}
<<\epsilon_{2}$, the particle number for each spin is given as: $N_{\uparrow}=Lq_{1\uparrow}/\pi$,
$N_{\uparrow}=Lq_{1\downarrow}/\pi$. As a result, $\bar{G}(0)=16N_{\uparrow}N_{\downarrow}
/(N_{\uparrow}+N_{\downarrow})^{2}$. Combining this result with Eq.(\ref{3}), it is direct
to obtain
\begin{eqnarray}
P=\sqrt{1-\frac{\bar{G}(0)}{4}}. \label{9}
\end{eqnarray}
If the superconductor is a normal $s$-wave superconductor, the ZBC in the strict
one-dimensional limit is given as \cite{SM}
\begin{eqnarray}
\tilde{G}(0)=\frac{e^{2}}{h}\frac{16\kappa^{2}q_{1\downarrow}q_{1\uparrow}}{(\kappa^{2}
+q_{1\uparrow}q_{1\downarrow}+\bar{Z}^{2})^{2}+
\bar{Z}^{2}(q_{1\downarrow}-q_{1\uparrow})^{2}},\label{10}
\end{eqnarray}
where $\kappa\simeq\sqrt{2m\mu_{s}}$, and $\bar{Z}=mZ/\hbar$. As $\tilde{G}(0)$ involves
parameters of all three parts: the FM, the superconductor, and the interface, the resolution
of SP from $\tilde{G}(0)$, if not possible, is very complicated \cite{I. I. Mazin1, T. Y. Chen}.
It is found that only in the clean limit and without mismatch of Fermi surface between
the FM and the superconductor, SP can be directly determined by $\tilde{G}(0)$ through the formula
\begin{eqnarray}
P=\left(1-\frac{\bar{\tilde{G}}(0)}{4}\right)^{\frac{1}{4}},\label{11}
\end{eqnarray}
where $\bar{\tilde{G}}(0)\equiv h\tilde{G}(0)/e^{2}$. Any one of the
two ideal conditions in real junctions is in fact hardly to satisfy.
All of these suggest that compared to the FM/$s$-wave superconductor
junction, the topological property of $\bar{G}(0)$ makes the resolution
of SP from the FM/TRITS junction much more direct and simplified,
simultaneously with an improvement of the precision.

When the number of subbands for spin-up and spin-down are both larger
than one, the simple formula (\ref{9}) is obviously no longer valid.
However, $\bar{G}(0)$ can still provide important information about
the SP. By defining a quantity as $\eta\equiv M/(\mu_{f}-\epsilon_{1})$,
which is the relative strength of the magnetization, it is direct to find
\begin{eqnarray}
\eta=\frac{4\sqrt{4-\bar{G}(0)}}{4+(4-\bar{G}(0))}. \label{12}
\end{eqnarray}
If we further define a quantity as $\lambda=\epsilon_{1}/(\mu_{f}-\epsilon_{1})$,
then the particle number for each spin can be expressed as: $N_{\uparrow}=(Lk_{f}/\pi)
\sum_{n}^{'}\sqrt{1+\eta-(n^{2}-1)\lambda}$, $N_{\downarrow}=(Lk_{f}/\pi)
\sum_{n}^{''}\sqrt{1-\eta-(n^{2}-1)\lambda}$, with $k_{f}=\sqrt{2m(\mu_{f}-\epsilon_{1})}$.
Therefore, if the value of $\lambda$ is known, the SP in fact can also be deduced from
$\bar{G}(0)$. As $\epsilon_{1}$ can easily be determined by measuring the width
of the ferromagnetic metal (if $m$ is known), the only residual challenge is to determine
$\mu_{f}$. However, if there are at least two subbands occupied by the minority spin electrons,
in fact we can easily determine $\lambda$ in the same setup by just tuning $\mu_{s}$.

\begin{figure}
\subfigure{\includegraphics[width=4cm, height=4cm]{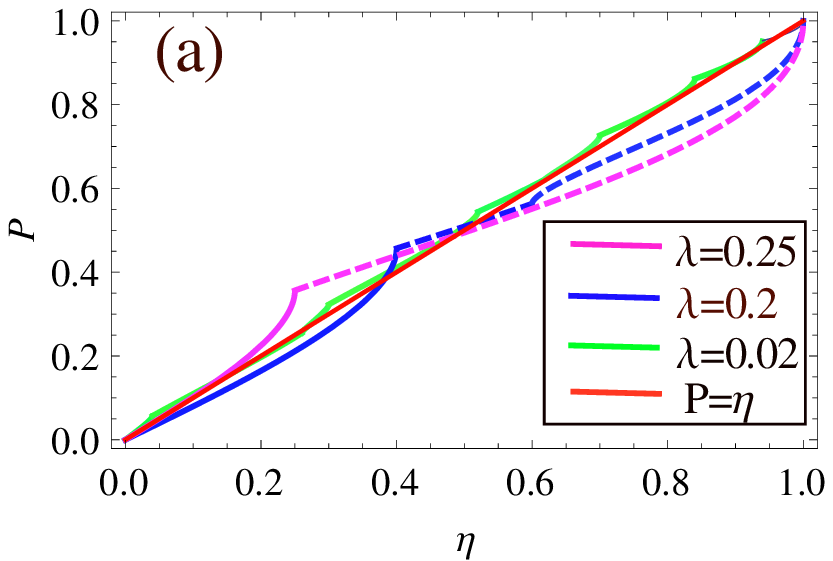}}
\subfigure{\includegraphics[width=4cm, height=4cm]{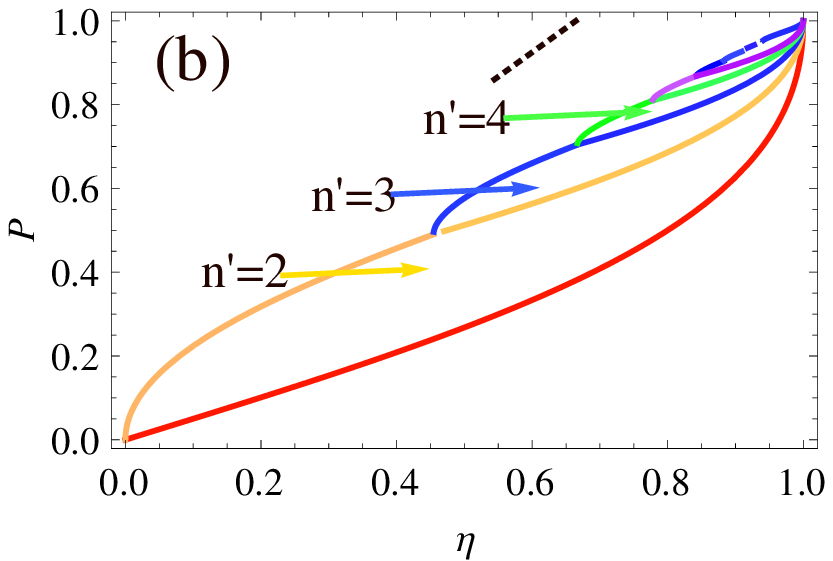}}
\caption{ (color online) Polarization-$vs$-magnetization. (a) The dashed parts of
the lines are where $\eta>1-3\lambda$, and as a result, $\lambda$ can not be determined
from Eq.(\ref{11}) due to $\delta \bar{G}(0)=0$. (b) With $n^{''}$ fixed to $1$, the region closed
by two neighbor lines corresponds to the possible value of the polarization
for certain $n^{'}$.}\label{fig1}
\end{figure}

As $\lambda$ enters into $N_{\uparrow}$ and $N_{\downarrow}$ through
terms with $n\geq2$, we can tune $\mu_{s}$ from the region $(\epsilon_{1},\epsilon_{2})$
to $(\epsilon_{2},\epsilon_{3})$. When $\mu_{s}$ goes across $\epsilon_{2}$, there is
a jump in ZBC, with $\delta \bar{G}(0)=16q_{2\uparrow}q_{2\downarrow}
/(q_{2\uparrow}+q_{2\downarrow})^{2}$, then a direct calculation gives
\begin{eqnarray}
\lambda=\frac{1}{3}\left[1-\sqrt{\left(\frac{\delta \bar{G}(0)}{8-\delta
\bar{G}(0)}\right)^{2}+\eta^{2}}\,\right].\label{13}
\end{eqnarray}
In Fig.\ref{fig1}(a), it is shown that when $\lambda<<1$ (many bands occupied),
the formula (\ref{13}) is almost valid in the whole region of $\eta$,  and $P$
is well approximated by $\eta$, $i.e.$, $P\simeq\eta$, this is a very useful
result for experiments.

If the number of occupied subbands is large, but the magnetization is so strong
that it makes $1>\eta>1-3\lambda$, then as $n^{''}=1$, $\delta \bar{G}(0)$ is always
equal to zero, the above approach to determine $\lambda$ breaks down. For this case,
as shown in Fig.\ref{fig1}(b), what we can be precisely determined is the upper limit
($P_{u}$) and lower limit ($P_{l}$) of the polarization. To character the uncertainty
of the polarization, we define a quantity as $\Delta\bar{P}=2(P_{u}-P_{l})/(P_{u}+P_{l})$.
When $n^{''}=1$, $n^{'}=2$, in the weak magnetization region, it is found that $\Delta\bar{P}$
can go beyond $100\%$. However, with $n^{'}$ increasing, $\Delta\bar{P}$ decreases very fast.
When $n^{'}\geq5$, we can take the boundary value $P_{u}$ or $P_{l}$, or $\eta$, as the
precise value of the SP. To determine the number of the subbands for the majority spin
electrons, we only need to detect the ZBC of the FM. If the FM is sufficiently clean to
guarantee $W<<L<l$, where $l$ is the mean free path, the quantized ZBC is given as
$G(0)=(n^{'}+n^{''})e^{2}/h$.

The three cases analyzed above exhaust all possibilities. For most cases, due to the
topological nature of $\bar{G}(0)$ and $\delta \bar{G}(0)$, the polarization can be easily and
precisely determined by this . Only in the parameter region: $n^{''}=1$, $2\leq n^{'}\leq4$,
the precision is not very good. If with the help of other measurements, both $\epsilon_{1}$
and $\mu_{f}$ are precisely determined, of course then $P$ can be directly and precisely
determined by the simple quantity $\bar{G}(0)$ for all cases.

{\it Magnetic proximity effect---} When a FM is in proximity
to a superconductor, the magnetization of the FM is equivalent to a magnetic field,
and it will penetrate into the superconductor and break Cooper pairs within the magnetic
penetration depth. However, this pair-breaking effect should not affect the validity of
the three formulae (\ref{9})(\ref{12})(\ref{13}), because the penetration is a local
behavior, it should not affect the topological property of the superconductor. In fact,
in real experiments, this pair-breaking effect can be avoided or greatly reduced by adding
a finite-thickness insulator between the FM and the superconductor. It is found that no matter
how thick the insulator is, when $\epsilon_{1}<\mu_{s}<\epsilon_{2}$, $\bar{G}(0)$
is always given by the formula (\ref{8}) \cite{SM}. Although $\bar{G}(0)$ does not
depend on the thickness of the insulator, the width of $\bar{G}(eV)$ will exponentially decrease
with the thickness. Therefore, for the sake of observing the peak and detecting its value, a proper
choice of the thickness is needed.

{\it Experimental realization---}  Compared to the TRI $p$-wave superconductor
with un-spin-polarized pairing type that belongs to the BDI class, a TRI $d$-wave
TS is in fact more experimentally realizable \cite{C. L. M. Wong, Z. Yan}.
Similar to the semiconductor-based proposal of TS \cite{Roman M. Lutchyn, Y. Oreg},
a TRI $d$-wave TS can also be realized by making a semiconductor wire with intrinsic
spin-orbit coupling in proximity to a $d$-wave superconductor \cite{C. L. M. Wong}.
Both materials are common in reality. As the TRI $d$-wave TS belongs to the DIII class,
it can only host at most one subband (without considering degeneracy) of nontrivial topology.
To determine both $\bar{G}(0)$ and $\lambda$, we can first tune $\mu_{s}$
to make only the lowest subband to be topological and obtain $\bar{G}(0)$,
and then tune $\mu_{s}$ to make only the second-lowest subband to be topological
and obtain $\bar{G}'(0)=\delta \bar{G}(0)$ \cite{SM}.

{\it Conclusions---}  The independence of $Z$ and the robustness against
magnetic proximity effect make the ZBC an observable that can easily and
directly determine the SP of a FM. This points out another potential
application of TS besides its well-known potential application in TQC.

{\it Acknowledgments.---}
This work was supported by NSFC Grant No.11275180.

\begin{widetext}
\section{Supplementary Materials}

\subsubsection{I. Ferromagnet/$s$-wave superconductor junction}
For a one-dimensional ferromagnet (FM)/$s$-wave superconductor junction, the Hamiltonians describing the
FM and the $s$-wave superconductor under the representation $\hat{\Psi}^{\dag}_{k}=(\hat{c}_{\uparrow,k}^{\dag},
\hat{c}_{\downarrow,-k},\hat{c}_{\downarrow,k}^{\dag},-\hat{c}_{\uparrow,-k})$  are given as
\begin{eqnarray}
&&\mathcal{H}_{F}(k)=[\frac{\hbar^{2}k^{2}}{2m}-\mu_{f}]\tau_{z}-M\tau_{0}\sigma_{z},\nonumber\\
&&\mathcal{H}_{S}(k)=[\frac{\hbar^{2}k^{2}}{2m}-\mu_{s}]\tau_{z}+\Delta\tau_{x}, \label{14}
\end{eqnarray}
respectively. When a spin-up electron with the Fermi energy is injected from the FM to the superconductor,
if we assume that the FM corresponds to the $x<0$ part, while the superconductor corresponds to the $x>0$
part, the general wave function in the FM is given as
\begin{eqnarray}
\psi_{f}(x)=\left(
                    \begin{array}{c}
                      1 \\
                      0 \\
                      0 \\
                      0 \\
                    \end{array}
                  \right)e^{iq_{\uparrow}x}+b_{\uparrow}\left(
                    \begin{array}{c}
                      1 \\
                      0 \\
                      0 \\
                      0 \\
                    \end{array}
                  \right)e^{-iq_{\uparrow}x}+a_{\downarrow}
                  \left(
                    \begin{array}{c}
                      0 \\
                      1 \\
                      0 \\
                      0 \\
                    \end{array}
                  \right)e^{iq_{\downarrow}x}+b_{\downarrow}\left(
                    \begin{array}{c}
                      0 \\
                      0 \\
                      1 \\
                      0 \\
                    \end{array}
                  \right)e^{-iq_{\downarrow}x}+a_{\uparrow}
                  \left(
                    \begin{array}{c}
                      0 \\
                      0 \\
                      0 \\
                      1 \\
                    \end{array}
                  \right)e^{iq_{\uparrow}x},\label{15}
\end{eqnarray}
where $q_{\uparrow}=\sqrt{2m(\mu_{f}+M)}$, $q_{\downarrow}=\sqrt{2m(\mu_{f}-M)}$. $b_{\uparrow(\downarrow)}$
denotes the amplitude that the injected electron is reflected as a spin-up (spin-down) electron, and
$a_{\uparrow(\downarrow)}$ denotes the amplitude that the injected electron is reflected as a spin-up
(spin-down) hole. The general wave function in the superconductor is given as
\begin{eqnarray}
\psi_{s}(x)=c_{1}\left(
                    \begin{array}{c}
                      i \\
                      1 \\
                      0 \\
                      0 \\
                    \end{array}
                  \right)e^{i\kappa x-\gamma x}+d_{1}\left(
                    \begin{array}{c}
                      -i \\
                      1 \\
                      0 \\
                      0 \\
                    \end{array}
                  \right)e^{-i\kappa x-\gamma x}+c_{2}\left(
                    \begin{array}{c}
                      0 \\
                      0 \\
                      i \\
                      1 \\
                    \end{array}
                  \right)e^{i\kappa x-\gamma x}+d_{2}\left(
                    \begin{array}{c}
                      0 \\
                      0 \\
                      -i \\
                      1 \\
                    \end{array}
                  \right)e^{-i\kappa x-\gamma x},\label{16}
\end{eqnarray}
where $\kappa=\left(\sqrt{2m\sqrt{\mu_{s}^{2}+\Delta^{2}}+2m\Delta}+\sqrt{2m\sqrt{\mu_{s}^{2}
+\Delta^{2}}-2m\Delta}\right)/2$, $\gamma=\left(\sqrt{2m\sqrt{\mu_{s}^{2}+\Delta^{2}}+2m\Delta}-
\sqrt{2m\sqrt{\mu_{s}^{2}+\Delta^{2}}-2m\Delta}\right)/2$. By matching the wave function at
$x=0$ according to the boundary conditions
\begin{eqnarray}
&&\psi_{f}(0)=\psi_{s}(0),\nonumber\\
&&v_{s}\psi_{s}(0^{+})-v_{f}\psi_{f}(0^{-})=-iZ\tau_{z}\sigma_{0}\psi_{s}(0),\label{17}
\end{eqnarray}
where $v_{s}=\partial_{k}\mathcal{H}_{s}/\hbar$, $v_{f}=\partial_{k}\mathcal{H}_{f}/\hbar$,
whose concrete expressions are given as
\begin{eqnarray}
v_{s}=v_{n}=\frac{\hbar}{m}\left(
        \begin{array}{cccc}
          k & 0 & 0 & 0 \\
          0 & -k & 0 & 0 \\
          0 & 0 & k & 0 \\
          0 & 0 & 0 & -k \\
        \end{array}
      \right)=\frac{-i\hbar}{m}\left(
        \begin{array}{cccc}
          \partial_{x} & 0 & 0 & 0 \\
          0 & -\partial_{x} & 0 & 0 \\
          0 & 0 & \partial_{x} & 0 \\
          0 & 0 & 0 & -\partial_{x} \\
        \end{array}
      \right),\label{18}
\end{eqnarray}
and $Z$ denotes the interface scattering potential, we obtain a series of algebraic relation
between the coefficients,
\begin{eqnarray}
&&1+b_{\uparrow}=i(c_{1}-d_{1}),\nonumber\\
&&a_{\downarrow}=c_{1}+d_{1},\nonumber\\
&&b_{\downarrow}=i(c_{2}-d_{2}),\nonumber\\
&&a_{\uparrow}=c_{2}+d_{2},\nonumber\\
&& ic_{1}(\kappa+i\gamma)+id_{1}(\kappa-i\gamma)-q_{\uparrow}(1-b_{\uparrow})=-i\bar{Z}(1+b_{\uparrow}),\nonumber\\
&&-c_{1}(\kappa+i\gamma)+d_{1}(\kappa-i\gamma)+q_{\downarrow}a_{\downarrow}=i\bar{Z}a_{\downarrow},\nonumber\\
&&ic_{2}(\kappa+i\gamma)+id_{2}(\kappa-i\gamma)+q_{\downarrow}b_{\downarrow}=-i\bar{Z}b_{\downarrow},\nonumber\\
&&-c_{2}(\kappa+i\gamma)+d_{2}(\kappa-i\gamma)+q_{\uparrow}a_{\uparrow}=i\bar{Z}a_{\uparrow},\label{19}
\end{eqnarray}
where $\bar{Z}=mZ/\hbar$. A direct calculation gives
\begin{eqnarray}
&&b_{\uparrow}=-\frac{\kappa^{2}-q_{\uparrow}q_{\downarrow}+(\gamma+\bar{Z})^{2}
+i(\gamma+\bar{Z})(q_{\uparrow}+q_{\downarrow})}{\kappa^{2}+q_{\uparrow}q_{\downarrow}+(\gamma+\bar{Z})^{2}
+i(\gamma+\bar{Z})(q_{\downarrow}-q_{\uparrow})},\nonumber\\
&&a_{\downarrow}=-\frac{2i\kappa q_{\uparrow}}{\kappa^{2}+q_{\uparrow}q_{\downarrow}+(\gamma+\bar{Z})^{2}
+i(\gamma+\bar{Z})(q_{\downarrow}-q_{\uparrow})},\nonumber\\
&&c_{1}=-\frac{iq_{\uparrow}(\kappa+q_{\downarrow}-i(\gamma+\bar{Z}))}{\kappa^{2}+q_{\uparrow}q_{\downarrow}
+(\gamma+\bar{Z})^{2}+i(\gamma+\bar{Z})(q_{\downarrow}-q_{\uparrow})},\nonumber\\
&&d_{1}=-\frac{iq_{\uparrow}(\kappa-q_{\downarrow}+i(\gamma+\bar{Z}))}{\kappa^{2}+q_{\uparrow}q_{\downarrow}
+(\gamma+\bar{Z})^{2}+i(\gamma+\bar{Z})(q_{\downarrow}-q_{\uparrow})},\nonumber\\
&&b_{\downarrow}=a_{\uparrow}=c_{2}=d_{2}=0.\label{20}
\end{eqnarray}
As the wave function in the superconductor decays with $x$ increasing, the non-vanishing two quantities $c_{1}$
and $d_{1}$ also have no contribution to the current. Therefore, the current conservation needs: $A_{\downarrow}+B_{\uparrow}=q_{\downarrow}|a_{\downarrow}|^{2}/q_{\uparrow}+|b_{\uparrow}|^{2}=1$,
which is easy to be verified. Then according to the Blonder-Tinkham-Klapwijk (BTK) formula \cite{G. E. Blonder1}, the zero-bias
conductance (ZBC) is given as
\begin{eqnarray}
G_{\uparrow}(0)=\frac{e^{2}}{h}(1+A_{\downarrow}-B_{\uparrow})=2\frac{e^{2}}{h}A_{\downarrow}.\label{21}
\end{eqnarray}
Similar procedures for the spin-down case will show that $G_{\downarrow}(0)=G_{\uparrow}(0)$,
and therefore, the measured ZBC is given as
\begin{eqnarray}
G(0)&=&G_{\downarrow}(0)+G_{\uparrow}(0)\nonumber\\
&=&\frac{e^{2}}{h}\frac{16\kappa^{2}q_{\downarrow}q_{\uparrow}}{(\kappa^{2}+q_{\uparrow}q_{\downarrow}+(\gamma+\bar{Z})^{2})^{2}+
(\gamma+\bar{Z})^{2}(q_{\downarrow}-q_{\uparrow})^{2}}.\label{22}
\end{eqnarray}
As generally $\mu_{s}>>\Delta$, $\kappa\simeq k_{F}=\sqrt{2m\mu_{s}}$, $\gamma\simeq\frac{\Delta}{\mu_{s}}k_{F}<<\kappa$,
$\gamma$ can be safely neglected, then $G(0)$ is simplified as
\begin{eqnarray}
G(0)=\frac{e^{2}}{h}\frac{16\kappa^{2}q_{\downarrow}q_{\uparrow}}{(\kappa^{2}+q_{\uparrow}q_{\downarrow}+\bar{Z}^{2})^{2}+
\bar{Z}^{2}(q_{\downarrow}-q_{\uparrow})^{2}}.\label{23}
\end{eqnarray}
$G(0)$ depends on the parameters of all three parts: the FM, the superconductor, and the interface.
It is obvious that the resolution of the spin polarization (SP) from $G(0)$ if not possible, is very
hard and complicated. In fact, only in the clean limit and without mismatch of the Fermi surface
between the FM and the superconductor, $i.e.$, $\bar{Z}=0$ and $\mu_{s}=\mu_{f}$, $G(0)$ can directly
determine the SP. Under the two idea conditions, $G(0)$ is simplified as
\begin{eqnarray}
G(0)=\frac{e^{2}}{h}\frac{16\sqrt{1-\bar{\eta}^{2}}}{(1+\sqrt{1-\bar{\eta}^{2}})^{2}},\label{24}
\end{eqnarray}
where $\bar{\eta}=M/\mu_{f}$. Then as $P=\frac{\sqrt{1+\bar{\eta}}-\sqrt{1-\bar{\eta}}}
{\sqrt{1+\bar{\eta}}+\sqrt{1-\bar{\eta}}}$, a direct calculation gives
\begin{eqnarray}
P=\left[1-\frac{\bar{G}(0)}{4}\right]^{\frac{1}{4}},\label{25}
\end{eqnarray}
where $\bar{G(0)}=hG(0)/e^{2}$.

\subsubsection{II. Ferromagnet/TRI $p$-wave superconductor with pairing type  $(|\uparrow\downarrow>+\downarrow\uparrow>)$ junction}

This junction is the case we have considered in the main text. The junction is
quasi-one-dimensional with width $W$, and the Hamiltonians describing
the FM and the superconductor under the representation $\hat{\Psi}^{\dag}_{nk}=(\hat{c}_{n\uparrow,k}^{\dag},
\hat{c}_{n\downarrow,-k},\hat{c}_{n\downarrow,k}^{\dag},\hat{c}_{n\uparrow,-k})$ are given as
\begin{eqnarray}
&&\mathcal{H}_{F}(k)=[\frac{\hbar^{2}k^{2}}{2m}
+\epsilon_{n}-\mu_{f}]\tau_{z}-M\tau_{0}\sigma_{z},\nonumber\\
&&\mathcal{H}_{pS}(k)=[\frac{\hbar^{2}k^{2}}{2m}
+\epsilon_{n}-\mu_{s}]\tau_{z}+\Delta(k)\tau_{x}, \label{26}
\end{eqnarray}
respectively, where $\epsilon_{n}=n^{2}\hbar^{2}\pi^{2}/(2mW^{2})$, $\Delta(k)=\Delta_{0}k$. With
the assumption $\epsilon_{1}<\mu_{s}<\epsilon_{2}$, when an spin-up electron with Fermi energy
and band index $n$ is injected from the FM ($x<0$) to the superconductor ($x>0$), the wave function
in the FM is given as
\begin{eqnarray}
\psi_{f,n}(x)=\left(
                    \begin{array}{c}
                      1 \\
                      0 \\
                      0 \\
                      0 \\
                    \end{array}
                  \right)e^{iq_{n\uparrow}x}+b_{n\uparrow}\left(
                    \begin{array}{c}
                      1 \\
                      0 \\
                      0 \\
                      0 \\
                    \end{array}
                  \right)e^{-iq_{n\uparrow}x}+a_{n\downarrow}
                  \left(
                    \begin{array}{c}
                      0 \\
                      1 \\
                      0 \\
                      0 \\
                    \end{array}
                  \right)e^{iq_{n\downarrow}x}+b_{n\downarrow}\left(
                    \begin{array}{c}
                      0 \\
                      0 \\
                      1 \\
                      0 \\
                    \end{array}
                  \right)e^{-iq_{n\downarrow}x}+a_{n\uparrow}
                  \left(
                    \begin{array}{c}
                      0 \\
                      0 \\
                      0 \\
                      1 \\
                    \end{array}
                  \right)e^{iq_{n\uparrow}x},\label{27}
\end{eqnarray}
where $q_{n\uparrow}=\sqrt{2m(\mu_{f}-\epsilon_{n}+M)}$,  $q_{n\downarrow}=\sqrt{2m(\mu_{f}-\epsilon_{n}-M)}$.
Note that due to the orthogonality of the wave functions in the confined direction, if there is no potential
depending on the coordinate in the confined direction, an electron with band index $n$ can only be reflected
as an electron or a hole with the same band index $n$. The wave function in the superconductor depends on the
band index, when $n=1$,
\begin{eqnarray}
\psi_{s,1}(x)=c_{11}\left(
                    \begin{array}{c}
                      i \\
                      1 \\
                      0 \\
                      0 \\
                    \end{array}
                  \right)e^{-k_{1+}x}+d_{11}\left(
                    \begin{array}{c}
                      i \\
                      1 \\
                      0 \\
                      0 \\
                    \end{array}
                  \right)e^{-k_{1-}x}+c_{12}\left(
                    \begin{array}{c}
                      0 \\
                      0 \\
                      i \\
                      1 \\
                    \end{array}
                  \right)e^{-k_{1+}x}+d_{12}\left(
                    \begin{array}{c}
                      0 \\
                      0 \\
                      i \\
                      1 \\
                    \end{array}
                  \right)e^{-k_{1-} x},\label{28}
\end{eqnarray}
and when $n\geq 2$,
\begin{eqnarray}
\psi_{s,n}(x)=c_{n1}\left(
                    \begin{array}{c}
                      i \\
                      1 \\
                      0 \\
                      0 \\
                    \end{array}
                  \right)e^{-k_{n+}x}+d_{n1}\left(
                    \begin{array}{c}
                      -i \\
                      1 \\
                      0 \\
                      0 \\
                    \end{array}
                  \right)e^{-k_{n-}x}+c_{n2}\left(
                    \begin{array}{c}
                      0 \\
                      0 \\
                      i \\
                      1 \\
                    \end{array}
                  \right)e^{-k_{n+}x}+d_{n2}\left(
                    \begin{array}{c}
                      0 \\
                      0 \\
                      -i \\
                      1 \\
                    \end{array}
                  \right)e^{-k_{n-} x},\label{29}
\end{eqnarray}
The concrete expressions of $k_{1\pm}$ depend on the relative magnitude of $(\mu_{s}-\epsilon_{1})$
and $\Delta_{0}$ \cite{Z. B. Yan1}. As they do not enter into the final results, their concrete
expressions do not matter, therefore, here we neglect the discussion on them. For $n\geq2$,
$k_{n\pm}=(\sqrt{m^{2}\Delta_{0}^{2}+2m(\epsilon_{n}-\mu_{s})}\pm m\Delta_{0})/\hbar$. $k_{n+}-k_{n-}=2m\Delta_{0}/\hbar=2\bar{\Delta}$ ($\bar{\Delta}\equiv m\Delta_{0}/\hbar$),
as we will see in the following, this is an important relation.

Now the boundary conditions at the interface $x=0$ are given as
\begin{eqnarray}
&&\psi_{f,n}(0)=\psi_{s,n}(0),\nonumber\\
&&v_{s,n}\psi_{s,n}(0^{+})-v_{f,n}\psi_{f}(0^{-})=-iZ\tau_{z}\sigma_{0}\psi_{s,n}(0),\label{30}
\end{eqnarray}
where $v_{f,n}$ takes the same form as $v_{f}$ in Eq.(\ref{18}), while $v_{s,n}$ now is given as
\begin{eqnarray}
v_{s,n}=\left(
        \begin{array}{cccc}
          \frac{-i\hbar}{m}\partial_{x} & \Delta_{0} & 0 & 0 \\
          \Delta_{0} & \frac{i\hbar}{m}\partial_{x} & 0 & 0 \\
          0 & 0 & \frac{-i\hbar}{m}\partial_{x} & \Delta_{0} \\
          0 & 0 & \Delta_{0} & \frac{i\hbar}{m}\partial_{x} \\
        \end{array}
      \right).\label{31}
\end{eqnarray}
By matching the wave functions according to the boundary conditions, we obtain a
series of algebraic relation between the coefficients. For $n=1$,
\begin{eqnarray}
&&1+b_{1\uparrow}=i(c_{11}+d_{11}),\nonumber\\
&&a_{1\downarrow}=c_{11}+d_{11},\nonumber\\
&&b_{1\downarrow}=i(c_{12}+d_{12}),\nonumber\\
&&a_{1\uparrow}=c_{12}+d_{12},\nonumber\\
&&c_{11}(-k_{1+}+\bar{\Delta})+d_{11}(-k_{1-}+\bar{\Delta})-q_{1\uparrow}(1-b_{1\uparrow})=\bar{Z}(c_{11}+d_{11}),\nonumber\\
&&ic_{11}(-k_{1+}+\bar{\Delta})+id_{11}(-k_{1-}+\bar{\Delta})+q_{1\downarrow}a_{1\downarrow}=i\bar{Z}(c_{11}+d_{11}),\nonumber\\
&&c_{12}(-k_{1+}+\bar{\Delta})+d_{12}(-k_{1-}+\bar{\Delta})+q_{1\downarrow}b_{1\downarrow}=\bar{Z}(c_{12}+d_{12}),\nonumber\\
&&ic_{12}(-k_{1+}+\bar{\Delta})+id_{12}(-k_{1-}+\bar{\Delta})+q_{1\uparrow}a_{1\uparrow}=i\bar{Z}(c_{12}+d_{12}).\label{32}
\end{eqnarray}
From the fifth and sixth line, it is easy to find $q_{1\uparrow}(1-b_{1\uparrow})=iq_{1\downarrow}a_{1\downarrow}$,
and from the first and second line, it is found $1+b_{1\uparrow}=ia_{1\downarrow}$. A combination of the two equations
directly gives
\begin{eqnarray}
&&b_{1\uparrow}=\frac{q_{1\uparrow}-q_{1\downarrow}}{q_{1\uparrow}+q_{1\downarrow}},\nonumber\\
&&a_{1\downarrow}=-\frac{2iq_{1\uparrow}}{q_{1\uparrow}+q_{1\downarrow}}.\label{33}
\end{eqnarray}
Since the waves corresponding to $c_{11}$ and $d_{11}$ have no contribution to
the current, we do not need to calculate out their concrete expressions.
Similarly, it is easy to find $b_{1\downarrow}=a_{1\uparrow}=0$. Therefore,
according to the BTK formula, the ZBC is given as
\begin{eqnarray}
G_{1\uparrow}(0)=\frac{e^{2}}{h}(1+A_{1\downarrow}-B_{\uparrow})=2\frac{e^{2}}{h}A_{1\downarrow}
=\frac{e^{2}}{h}\frac{8q_{1\uparrow}q_{1\downarrow}}{(q_{1\uparrow}+q_{1\downarrow})^{2}},\label{34}
\end{eqnarray}
Similarly analysis for spin-down case finds $G_{1\downarrow}(0)=G_{1\uparrow}(0)$, and therefore,
\begin{eqnarray}
G_{1}(0)=G_{1\downarrow}(0)+G_{1\uparrow}(0)
=\frac{e^{2}}{h}\frac{16q_{1\uparrow}q_{1\downarrow}}{(q_{1\uparrow}+q_{1\downarrow})^{2}}.\label{35}
\end{eqnarray}
For $n\geq2$,
\begin{eqnarray}
&&1+b_{n\uparrow}=i(c_{n1}-d_{n1}),\nonumber\\
&&a_{n\downarrow}=c_{n1}+d_{n1},\nonumber\\
&&b_{n\downarrow}=i(c_{n2}-d_{n2}),\nonumber\\
&&a_{n\uparrow}=c_{n2}+d_{n2},\nonumber\\
&&c_{n1}(-k_{n+}+\bar{\Delta})-d_{n1}(k_{n-}+\bar{\Delta})-q_{n\uparrow}(1-b_{1\uparrow})=\bar{Z}(c_{n1}-d_{n1}),\nonumber\\
&&ic_{n1}(-k_{n+}+\bar{\Delta})-id_{n1}(k_{n-}+\bar{\Delta})+q_{n\downarrow}a_{1\downarrow}=i\bar{Z}(c_{n1}+d_{n1}),\nonumber\\
&&c_{n2}(-k_{n+}+\bar{\Delta})-d_{n2}(k_{n-}+\bar{\Delta})+q_{n\downarrow}b_{1\downarrow}=\bar{Z}(c_{n2}-d_{n2}),\nonumber\\
&&ic_{n2}(-k_{n+}+\bar{\Delta})-id_{n2}(k_{n-}+\bar{\Delta})+q_{n\uparrow}a_{1\uparrow}=i\bar{Z}(c_{n2}+d_{n2}).\label{36}
\end{eqnarray}
From the second line and the sixth line, we can obtain
\begin{eqnarray}
c_{n1}=\frac{k_{n-}+\bar{\Delta}+Z+iq_{n\downarrow}}{-k_{n+}+\bar{\Delta}-Z-iq_{n\downarrow}}d_{n1},\nonumber
\end{eqnarray}
As $k_{n+}-k_{n-}=2\bar{\Delta}$, which is equivalent to $-k_{n+}+\bar{\Delta}=-(k_{n-}+\bar{\Delta})$,
therefore, $c_{n1}=-d_{n1}$. Consequently, $a_{n\downarrow}=c_{n1}+d_{n1}=0$. Similarly, a combination of
the fourth line and the eighth line shows $c_{n2}=-d_{n2}$, and therefore, $a_{n\uparrow}=c_{n2}+d_{n2}=0$.
The remaining equations are reduced as
\begin{eqnarray}
&&1+b_{n\uparrow}=2ic_{n1},\nonumber\\
&&b_{n\downarrow}=2ic_{n2},\nonumber\\
&&q_{n\uparrow}(1-b_{1\uparrow})=-2\bar{Z}c_{n1},\nonumber\\
&&q_{n\downarrow}b_{1\downarrow}=2\bar{Z}c_{n2},\nonumber
\end{eqnarray}
it is easy to obtain $b_{n\downarrow}=c_{n2}=0$, and
\begin{eqnarray}
&&b_{n\uparrow}=\frac{q_{n\uparrow}-i\bar{Z}}{q_{n\uparrow}+i\bar{Z}},\nonumber\\
&&c_{n1}=-d_{n1}=-\frac{iq_{n\uparrow}}{q_{n\uparrow}+i\bar{Z}}.\label{37}
\end{eqnarray}
The electron is completely reflected as itself, therefore $G_{n\uparrow}(0)=0$. Similar analysis
for spin-down case also shows $G_{n\downarrow}(0)=0$, consequently,
\begin{eqnarray}
G_{n}(0)=G_{n\uparrow}(0)+G_{n\downarrow}(0)=0,\label{38}
\end{eqnarray}
Therefore, when $\epsilon_{1}<\mu_{s}<\epsilon_{2}$, the total ZBC is given as
\begin{eqnarray}
G(0)=\sum_{n}G_{n}(0)=G_{1}(0)
=\frac{e^{2}}{h}\frac{16q_{1\uparrow}q_{1\downarrow}}{(q_{1\uparrow}+q_{1\downarrow})^{2}}.\label{39}
\end{eqnarray}

\subsection{III: Ferromagnet/insulator/TRI $p$-wave superconductor with pairing type  $(|\uparrow\downarrow>+|\downarrow\uparrow>)$ junction}

When an insulator ($-d<x<0$) is inserted between the FM ($x<-d$) and the superconductor ($x>0$),
the wave functions corresponding to a left-injected spin-up electron with Fermi energy and band
index $n$ in the FM and the superconductor are also given as $\psi_{f,n}(x)$ and $\psi_{s,n}(x)$,
respectively, and the wave function in the insulator is given as
\begin{eqnarray}
\psi_{I,n}(x)&=&c_{n1}\left(\begin{array}{c}
                      1 \\
                      0 \\
                      0 \\
                      0 \\
                    \end{array}\right)e^{q_{I,n}x}+d_{n1}\left(\begin{array}{c}
                      1 \\
                      0 \\
                      0 \\
                      0 \\
                    \end{array}\right)e^{-q_{I,n}x}+c_{n2}\left(\begin{array}{c}
                      0 \\
                      1 \\
                      0 \\
                      0 \\
                    \end{array}\right)e^{q_{I,n}x}+d_{n2}\left(\begin{array}{c}
                      0 \\
                      1 \\
                      0 \\
                      0 \\
                    \end{array}\right)e^{-q_{I,n}x} \nonumber\\
&+&c_{n3}\left(\begin{array}{c}
                      0 \\
                      0 \\
                      1 \\
                      0 \\
                    \end{array}\right)e^{q_{I,n}x}+d_{n3}\left(\begin{array}{c}
                      0 \\
                      0 \\
                      1 \\
                      0 \\
                    \end{array}\right)e^{-q_{I,n}x}+c_{n4}\left(\begin{array}{c}
                      0 \\
                      0 \\
                      0 \\
                      1 \\
                    \end{array}\right)e^{q_{I,n}x}+d_{n4}\left(\begin{array}{c}
                      0 \\
                      0 \\
                      0 \\
                      1 \\
                    \end{array}\right)e^{-q_{I,n}x},\label{40}
\end{eqnarray}
where $q_{I,n}=\sqrt{2m(\mu_{I}+\epsilon_{n})}$. The boundary conditions are given as
\begin{eqnarray}
&&\psi_{f,n}(-d)=\psi_{I,n}(-d),\nonumber\\
&&\psi_{I,n}(0)=\psi_{s,n}(0),\nonumber\\
&&v_{i,n}\psi_{I,n}(-d^{+})-v_{f,n}\psi_{f,n}(-d^{-})=-iZ_{1}\tau_{z}\sigma_{0}\psi_{I,n}(-d),\nonumber\\
&&v_{s,n}\psi_{s,n}(0^{+})-v_{i,n}\psi_{I,n}(0^{-})=-iZ_{2}\tau_{z}\sigma_{0}\psi_{s,n}(0),\label{41}
\end{eqnarray}
where $v_{s,n}$ is also given by Eq.(\ref{31}), $v_{f,n}$ is given by Eq.(\ref{18}), and $v_{i,n}=v_{f,n}$.
Then a direct calculation shows: when $n=1$,
\begin{eqnarray}
&&b_{\uparrow}=\frac{q_{1\uparrow}-q_{1\downarrow}}{q_{1\uparrow}+q_{1\downarrow}}e^{-2iq_{1\uparrow}d},\nonumber\\
&&a_{\downarrow}=-i\frac{2q_{1\uparrow}}{q_{1\uparrow}+q_{1\downarrow}}e^{i(q_{1\downarrow}-q_{1\uparrow})d},\nonumber\\
&&b_{\downarrow}=a_{\uparrow}=0, \label{42}
\end{eqnarray}
and when $n=2$,
\begin{eqnarray}
&&b_{\uparrow}=\left(\frac{q_{1}[(q_{I}-\bar{Z}_{2})e^{-q_{I}d}+(q_{I}+\bar{Z}_{2})e^{q_{I}d}]
-i[(q_{I}+\bar{Z}_{1})(q_{I}+\bar{Z}_{2})e^{q_{I}d}-(q_{I}-\bar{Z}_{1})(q_{I}-\bar{Z}_{2})e^{-q_{I}d}]}
{q_{1}[(q_{I}-\bar{Z}_{2})e^{-q_{I}d}+(q_{I}+\bar{Z}_{2})e^{q_{I}d}]
+i[(q_{I}+\bar{Z}_{1})(q_{I}+\bar{Z}_{2})e^{q_{I}d}-(q_{I}-\bar{Z}_{1})(q_{I}-\bar{Z}_{2})e^{-q_{I}d}]}\right)
e^{-2iq_{n\uparrow}d},\nonumber\\
&&a_{\downarrow}=b_{\downarrow}=a_{\uparrow}=0,\label{43}
\end{eqnarray}
(other parameters are not given since they are not important here) it is direct to see that the thickness
of the insulator, $d$, only affects the phases of the coefficients, it does not affect the magnitude of the
Andreev reflection and normal reflection coefficients. Therefore, no matter how large $d$ is, $G_{1}(0)$,
$G_{n\geq2}$, and $G(0)$ are always given by Eq.(\ref{35}), Eq.(\ref{38}) and  Eq.(\ref{39}), respectively.

\subsubsection{IV. Ferromagnet/TRI $d$-wave superconductor with pairing type  $(|\uparrow\downarrow>-|\downarrow\uparrow>)$ junction}

When the TRI TS is a one-dimensional spin-orbit coupled $d$-wave superconductor, the Hamiltonian
under the representation $\hat{\Psi}_{k}=(\hat{c}_{k\uparrow}^{\dag},
\hat{c}_{k\downarrow}^{\dag},\hat{c}_{-k\uparrow},\hat{c}_{-k\downarrow})$ is given as
\cite{C. L. M. Wong1, Z. Yan1},
\begin{eqnarray}
\mathcal{H}_{dS}(k)=(-t\cos(k)-\mu_{s})\sigma_{0}\tau_{z}+\alpha\sin k \sigma_{y}\tau_{z}+ \Delta \cos k \sigma_{y}\tau_{y},\label{44}
\end{eqnarray}
this is a tight-binding form. As $\sigma_{y}\mathcal{H}_{dS}(k)\sigma_{y}=\mathcal{H}_{dS}^{*}(-k)$,
$\tau_{x}\mathcal{H}_{dS}(k)\tau_{x}=-\mathcal{H}_{dS}^{*}(-k)$, $\sigma_{y}\tau_{x}\mathcal{H}_{dS}(k)
\tau_{x}\sigma_{y}=-\mathcal{H}_{dS}(k)$, the Hamiltonian belongs to the DIII class that is characterized
by a $Z_{2}$ invariant \cite{A. P. Schnyder1, A. Y. Kitaev1}. The energy spectrum is given as
\begin{eqnarray}
E_{k}=\pm\sqrt{(-t\cos k-\mu_{s}\pm\alpha \sin k)^{2}+\Delta^{2}\cos^{2} k}.\label{45}
\end{eqnarray}
As the term $-t\cos k$ does not affect the gap closing condition, in fact,
we can neglect it. Then the energy spectrum is simplified to
\begin{eqnarray}
E_{k}=\pm\sqrt{(\mu_{s}\pm\alpha \sin k)^{2}+\Delta^{2}\cos^{2} k}.\label{46}
\end{eqnarray}
Without loss of generality, we assume $\mu_{s}>0, \alpha>0$, then the gap is closed
at $k=\pm\pi/2$ only when $\mu_{s}=\alpha$. Assuming the parameters are in the
topological regime $\mu_{s}<\alpha$ and at the neighborhood of the gap closing point,
then by a low-energy expansion at $k=\pi/2$, we obtain the low-energy spectrum
\begin{eqnarray}
E_{k}=\pm\sqrt{({\frac{k^{2}}{2\tilde{m}}-\tilde{\mu})^{2}+\Delta^{2} k^{2}}}, \label{47}
\end{eqnarray}
where $\tilde{\mu_{s}}=\alpha-\mu_{s}$ and $\tilde{m}=1/\alpha$. The other expansion
at $k=-\pi/2$ takes the same form and plays the role of a time-reversal partner. From
Eq.(\ref{47}), it is direct to see that the spin-orbit coupled $d$-wave superconductor
has been mapped to a $p$-wave superconductor.  For the spin-orbit coupled $d$-wave
superconductor, the topological critical point is $|\alpha|=|\mu_{s}|$ \cite{C. L. M. Wong1, Z. Yan1},
which is just equivalent to $\tilde{\mu_{s}}=0$, this verifies that the mapping is indeed
valid. Therefore, in the quasi-one-dimensional case, if $\epsilon_{1}<\mu_{s}<\epsilon_{2}$
and $(\mu_{s}-\epsilon_{1})<\alpha<(\epsilon_{2}-\mu_{s})$, the spin-orbit coupled
$d$-wave superconductor is a TS with only the lowest subband (without considering degeneracy)
of nontrivial topology. Then as the $d$-wave pairing is also an un-spin-polarized type,
the ZBC of the FM/TRI $d$-wave superconductor junction is also given as $G(0)=\frac{e^{2}}{h}\frac{16q_{n}q_{\bar{n}}}{(q_{n}+q_{\bar{n}})^{2}}$
(we have verified this result by a direct calculation). Similar to the TRI $p$-wave superconductor
considered in the main text, if there are at least two subbands occupied by the minority spin
electrons, by tuning the parameter to satisfy: $\mu_{s}>\epsilon_{2}$, $(\mu_{s}-\epsilon_{2})
<\alpha<(\mu_{s}-\epsilon_{1})$, then only the subbands with index $n=2$ become topological while the
original topological subbands with index $n=1$ turn to be topologically trivial, and we can obtain
another ZBC $G'(0)$ which is equivalent to $\delta G(0)$ in the main text. Therefore, in this case,
a combination of $G(0)$ and a non-zero $G'(0)$ can also determine the value of $\lambda$. All
these results suggest that all TRITSs with un-spin-polarized pairing type are equivalent in determining
the SP of a FM.

\end{widetext}

\end{document}